\documentclass[letterpaper,12pt]{article}
\usepackage{tabularx} 
\usepackage{amsmath}  
\usepackage[margin=1in,letterpaper]{geometry} 
\usepackage{cite} 
\usepackage[final]{hyperref} 
\usepackage{authblk}
\usepackage[pdftex]{graphicx}
\hypersetup{
	colorlinks=true,       
	linkcolor=blue,        
	citecolor=blue,        
	filecolor=magenta,     
	urlcolor=blue         
}

\usepackage{mathtools}

\newcommand{\Tabref}[1]{Table \ref{#1}}
\newcommand{\Figref}[1]{Figure \ref{#1}}
\newcommand{\Equref}[1]{Equation \ref{#1}}

\begin{document}

\title{Study of the $h \gamma Z$ coupling at the ILC}

\author{Yumi Aoki$^{1}{}^{\dagger}$, Keisuke Fujii$^{2}$, Junping Tian$^{3}$\\
on behalf of the ILD concept group}

\affil{SOKENDAI$^{1}$, KEK$^{2}$, University of Tokyo$^{3}$\\
$\dagger$ yumia@post.kek.jp
}
\date{}

\maketitle

\begin{abstract}
We studied the $e^+e^- \to h \gamma $ process at the International Linear Collider (ILC)~\cite{Aus0} at $\sqrt{s}=250$ GeV, based on the full detector simulation of
the International Large Detector (ILD). This process is loop-induced in the Standard Model (SM) and is sensitive to new physics which alters $h \gamma \gamma$ or $h \gamma Z$ coupling. We performed the analysis by employing the leading signal channels with $h \to b \bar{b}$ and $h \to WW^*$ and including full SM background processes. The results are obtained for two scenarios of beam polarisations each with an integrated luminosity of 900 fb$^{-1}$. We found the expected significance of the SM signal is 0.40$\sigma$ for $P(e^-,e^+)=(-0.8,+0.3)$ (the left-handed polarisation), and 0.06$\sigma$ for $P(e^-,e^+)=(+0.8,-0.3)$ (the right-handed polarisation). The bounds on new physics effects are reported as the 95\% C.L.  upper limit for the cross-section of $e^+e^- \to h \gamma$: $\sigma_{h\gamma}^L <$ 1.8 fb and $\sigma_{h\gamma}^R <$ 0.5 fb respectively for left- and right-handed polarisations. The constraints on effective $h\gamma Z$ couplings are to be further studied.
\footnote{
Talk presented at the International Workshop on Future Linear Colliders (LCWS2021), 15-18 March 2021. C21-03-15.1.}

\end{abstract}

\section{Introduction}

With the discovery of the Higgs boson in 2012, the Standard Model (SM) has reached a point of completion. However, there are still phenomena that cannot be explained by the SM, and finding new physics beyond the SM (BSM) is an urgent task. The Higgs boson is a key to unraveling new physics.

The ILC is a proposed Higgs factory for discovering new physics. It is an electron-positron linear collider. The advantages of the ILC include its capability of delivering polarized beams as well as its energy extendability up to $\sqrt{s}\sim1$ TeV.

\begin{figure}[htbp] 
\begin{center}
\includegraphics[width=0.5\linewidth]{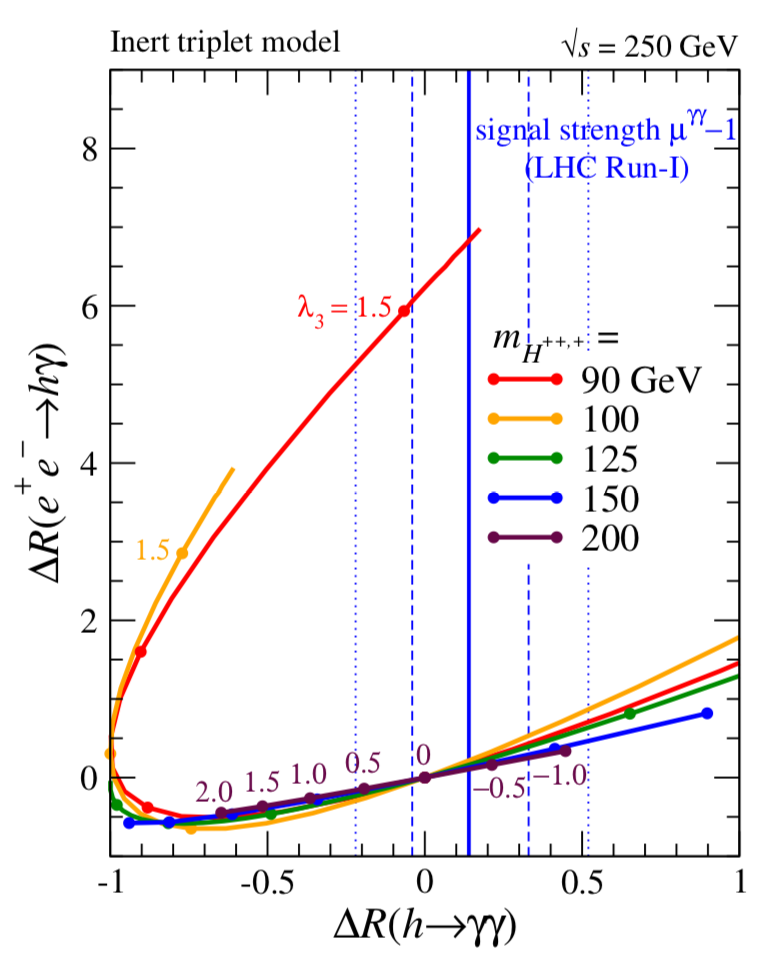}
\caption{The relative deviations from the Standard Model values for the $e^+ e^- \to h \gamma $ cross-section versus the $h \xrightarrow{} \gamma\gamma$ partial width in the Inert Triplet Model ~\cite{Aus2}.}
\label{fig:1}
\end{center}
\end{figure}

The main motivation of this study is to probe new physics via $h\gamma \gamma$ and $h \gamma Z$ couplings using the process $e^+e^- \to h \gamma$. In the SM these couplings are only loop-induced, thus may receive relatively large modifications from BSM contributions. One example of new physics is the Inert Triplet Model~\cite{Aus2} in which there are charged Higgs bosons contributing to the loop in the $h \gamma Z$ coupling.
\Figref{fig:1} shows the relative deviations from the SM values for the $e^+e^- \to h \gamma $ cross-section versus the $h \to \gamma \gamma$ partial width. 
Depending on the masses of charged Higgs bosons and other parameters of the model, the deviations can be as large as O(1). 

This paper is organized as follows. Chapter 2 describes the theoretical framework and experimental methods. Chapter 3 describes the simulation method. Chapter 4 describes the event selection, and Chapter 5 describes the results. In Chapter 6, we will explain about the uncertainty of this simulation. Finally, we will summarize our result and show the future study in Chapter 7.

\section{Theoretical Framework and Experimental Method}

\begin{figure}[htbp] 
        
        \centering \includegraphics[width=0.9\columnwidth]{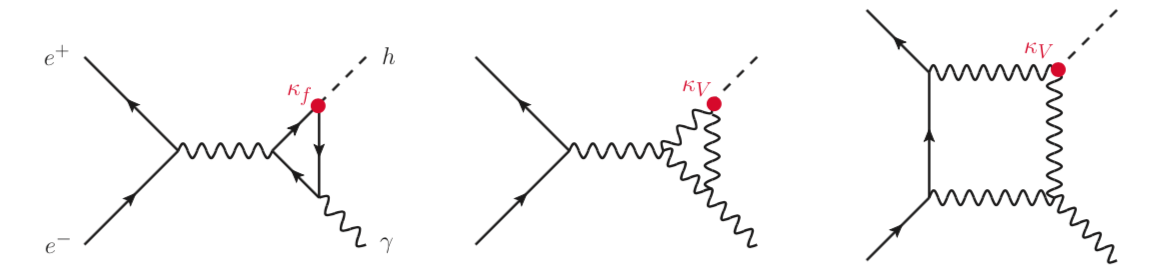}
        \caption{
                \label{fig:7} 
              The loop-induced Feynman diagrams in the SM for $e^+e^- \to h \gamma $~\cite{Aus2}. An exchanged particle of the left diagram is top quark, that of the center diagram is W boson, the last diagram is box diagram.}
\end{figure}

\begin{figure}[htbp] 
        
        \centering \includegraphics[width=0.6\columnwidth]{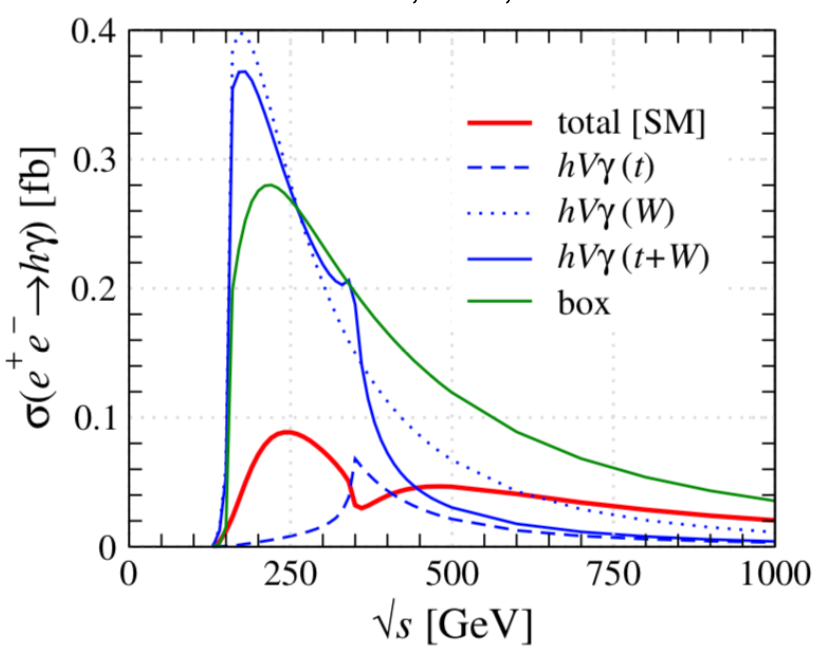}
        \caption{
                \label{fig:20} 
              The contributions from individual diagrams of \Figref{fig:7}. The long dashed line is for the diagram with top-quark in the loop (left); the dot-dashed line is for the triangle diagram with $W$ in the loop (middle); the green line is for the box diagram with $W$ in the loop (right); the blue solid line is for the sum of left and middle diagrams; the red line is for the total contribution.}
\end{figure}

We will first describe the theoretical framework. The main Feynman diagrams for $e^+e^- \to h \gamma $ are shown in \Figref{fig:7}. The contributions from the individual diagrams are shown in \Figref{fig:20}. It can be clearly seen that there is a large destructive interference in this process.
The cross-sections in the SM are shown in \Tabref{tbl:1} for different values of beam polarisations, from where we can expect that the measurement of this process is going to be challenging due to its small cross-sections well below 1 fb.

\begin{table}[htbp]
\begin{center}
\caption{SM cross-sections for $e^+e^- \to h \gamma $ for different beam polarization ($\sqrt{s}$  = 250 GeV ).}
\label{tbl:1} 
\begin{tabular}{cccc} 
\hline
\multicolumn{1}{c}{$P_{e^-}$ } & \multicolumn{1}{c}{$P_{e^+}$} & \multicolumn{1}{c}{$\sigma_{SM}$[fb]}  \\
\hline
-100$\%$ & +100$\%$ &  0.35\\
+100$\%$ & -100$\%$	& 0.016\\
-80$\%$ & +30$\%$ & 0.20\\

\hline
\end{tabular}
\end{center}
\end{table}

\begin{figure}[htbp] 
        
        \centering \includegraphics[width=0.9\columnwidth]{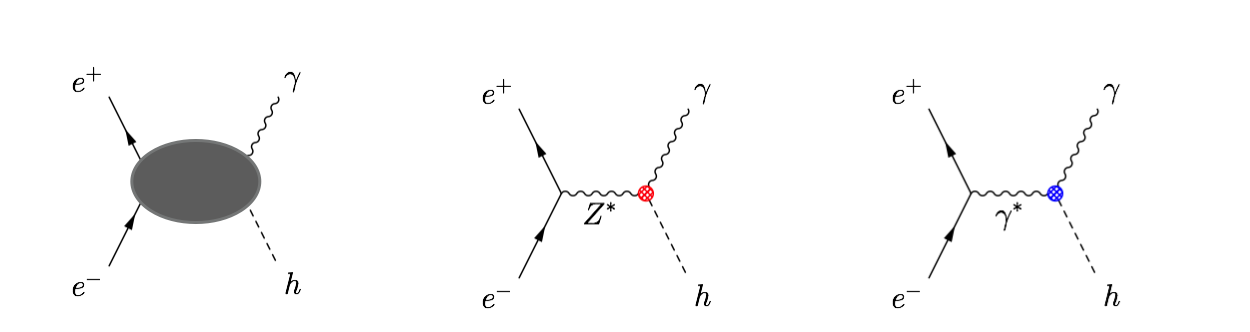}
        \caption{
                \label{fig:5} 
              Diagrams arising from each of the three terms of \Equref{lsm}, respectively. }
\end{figure}

\clearpage
Equation \ref{lsm} shows the effective field theory (EFT) Lagrangian which includes new physics contribution to $e^+ e^- \xrightarrow{} h \gamma $ in a model-independent way,
\begin{eqnarray}
{\cal{L}} _ {h \gamma } = {\cal{L}} _ { \mathrm { SM } } + \frac { \zeta _ { A Z } } { v } A _ { \mu \nu } Z ^ { \mu \nu } h + \frac { \zeta _ { A } } { 2 v } A _ { \mu \nu } A ^ { \mu \nu } h, 
\label{lsm}
\end{eqnarray}
where $A_{\mu \nu}$, $Z_{\mu \nu}$ are field strength tensors for photon and $Z$, $v$ is the vacuum expectation value. The first term stands for SM Lagrangian, the second term is effective $h \gamma Z$ interaction from BSM, and the third term is effective $h \gamma \gamma$ interaction from BSM. The diagram of each term is shown in \Figref{fig:5}.

Coefficient $\zeta_{A}$ can be determined precisely by the measurement of $h \to \gamma \gamma$ partial width. Coefficient $\zeta_{AZ}$ in principle can be also determined by the measurement of $h \to \gamma Z$ partial width which however is expected to be much less precise than $h \to \gamma \gamma$. Thus we are motivated to determine $\zeta_{AZ}$ in a complementary way, that is to use a measurement of the cross-section of $e^+ e^- \xrightarrow{} h \gamma $ at the ILC. \Equref{Equ:11} and \Equref{Equ:12}~\cite{crosssection} show the computation of the cross-section of $e^+ e^- \xrightarrow{} h \gamma $ (normalized by its SM value) in terms of $\zeta_{AZ}$ and $\zeta_A$ for 100\% left- and right-handed polarization respectively:
\begin{eqnarray}
\frac { \sigma _ { h\gamma  }^L } { \sigma _ { S M } } = 1 - 201 \zeta _ { A } - 273 \zeta _ { A Z }.
\label{Equ:11}
\end{eqnarray}
\begin{eqnarray}
\frac { \sigma _ { h \gamma  }^R } { \sigma _ { S M } } = 1 + 492 \zeta _ { A } - 311 \zeta _ { A Z }.
\label{Equ:12}
\end{eqnarray}

\section{Simulation framework}
\begin{figure}[htbp] 
        \centering \includegraphics[width=0.9\columnwidth]{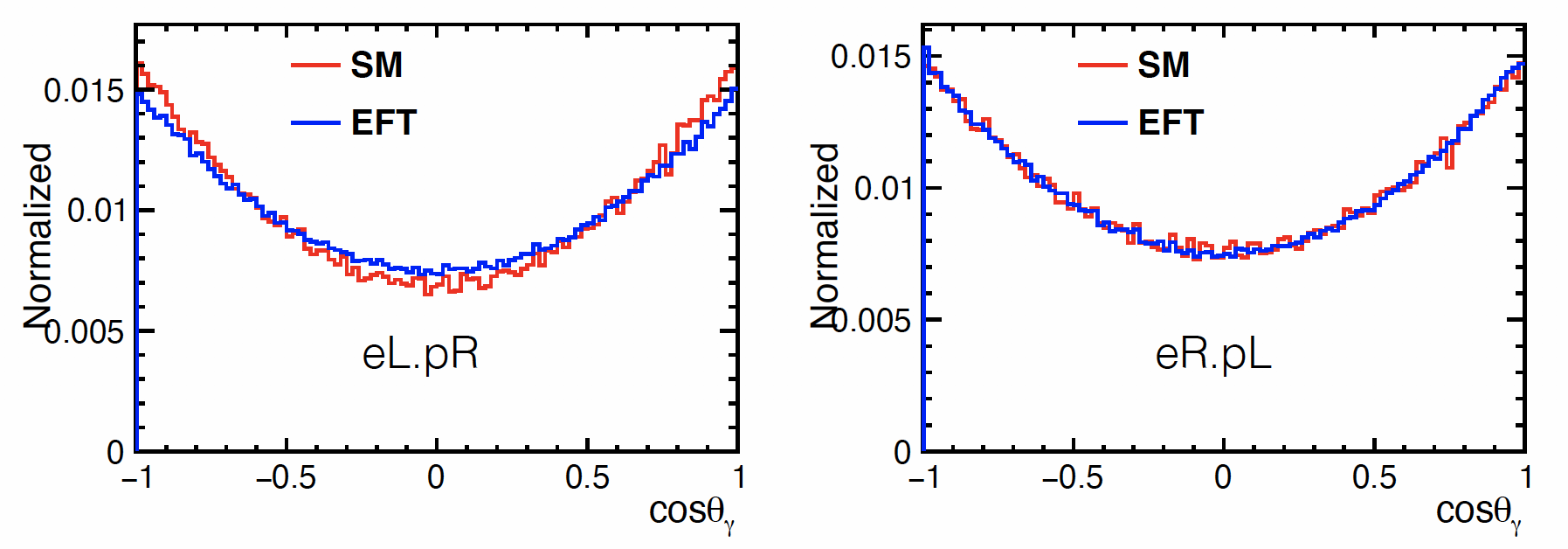}
        \caption{
        \label{fig:angular} 
        The angular distribution of each polarization (calculated by J. Tian). Photon polar angle distributions from $e^+e^- \to h \gamma$, using only EFT term of the Lagrangian (blue) and the full implementation of the \Equref{lsm} (red).}
\end{figure}

Next, we describe the simulation framework. First, we generate the signal events using Physsim~\cite{Aus5}. The implementation of signal generator has been updated from our earlier study~\cite{previousproceedings}: only the EFT term for $\zeta_{AZ}$ was employed for matrix element calculation in the earlier study now the SM full 1-loop contribution is properer implemented. It turns out that the angular distributions of photon production angle are not significantly changed, as shown in \Figref{fig:angular}. For background, we use the events which have been produced for the ILD DBD study~\cite{Aus4} using Whizard~\cite{Aus6}. 
Full SM $e^+e^-\to$ 2-fermion (2f) mainly $\gamma Z \to \gamma (\text{f f})$ and 4-fermion (4f) mainly $Z^+Z^-/W^+W^- \to 4\text{f} $ background events are included in this analysis.

All generated events are fully simulated with Geant4~\cite{Aus8} for the ILD DBD model using Mokka~\cite{Aus7}. The simulated events are then reconstructed use Marlin in iLCSoft~\cite{Aus9}, where particle flow analysis (PFA) is performed with PandoraPFA~\cite{Aus10} and flavor tagging is done with LCFI+~\cite{Aus11}. The analysis is carried out at $\sqrt{s}$ = 250 GeV. We assume an integrated luminosity of 900 fb$^{-1}$ for both $P(e^-,e^+)=(-0.8,+0.3)$ (left-handed case) and $P(e^-,e^+)=(+0.8,-0.3)$ (right-handed case).

\section{Event selection}
Here we describe the procedure of signal event selection and background event suppression.
The signal production process is $e^+e^-\to h\gamma $. We focus on the leading Higgs decay channels: $h\to b\bar{b}$ and $h\to WW^*$ semi-leptonic final states. Both signal channels are characterized by the presence of an isolated photon. First, for every event, it is necessary to find at least one photon with an energy greater than 50 GeV in every event. The photon identification algorithm is suppliered by PandoraPFA. The most energetic photon, if more than one is found, is then selected as the signal photon. The remaining particles in every event other than the signal photon undergo the following channel-specific event selections.

\subsection{$h \to b\bar{b}$ channel}
The characteristic of the signal events in this channel is one monochromatic energetic photon with an energy of about 93 GeV and two b-jets of which invariant mass is consistent with the Higgs mass of 125 GeV. Thus the particles other than the signal photon identified above are clustered into 2 jets using the Durham jet clustering algorithm~\cite{Aus13}. The 2 jets are both flavor-tagged using the algorithm in LCFI+. 

\begin{figure}[htbp]
\begin{center}
	\begin{tabular}{c}
	\begin{minipage}{0.5\hsize}
	\begin{flushleft}
	\includegraphics[width=83mm,clip]{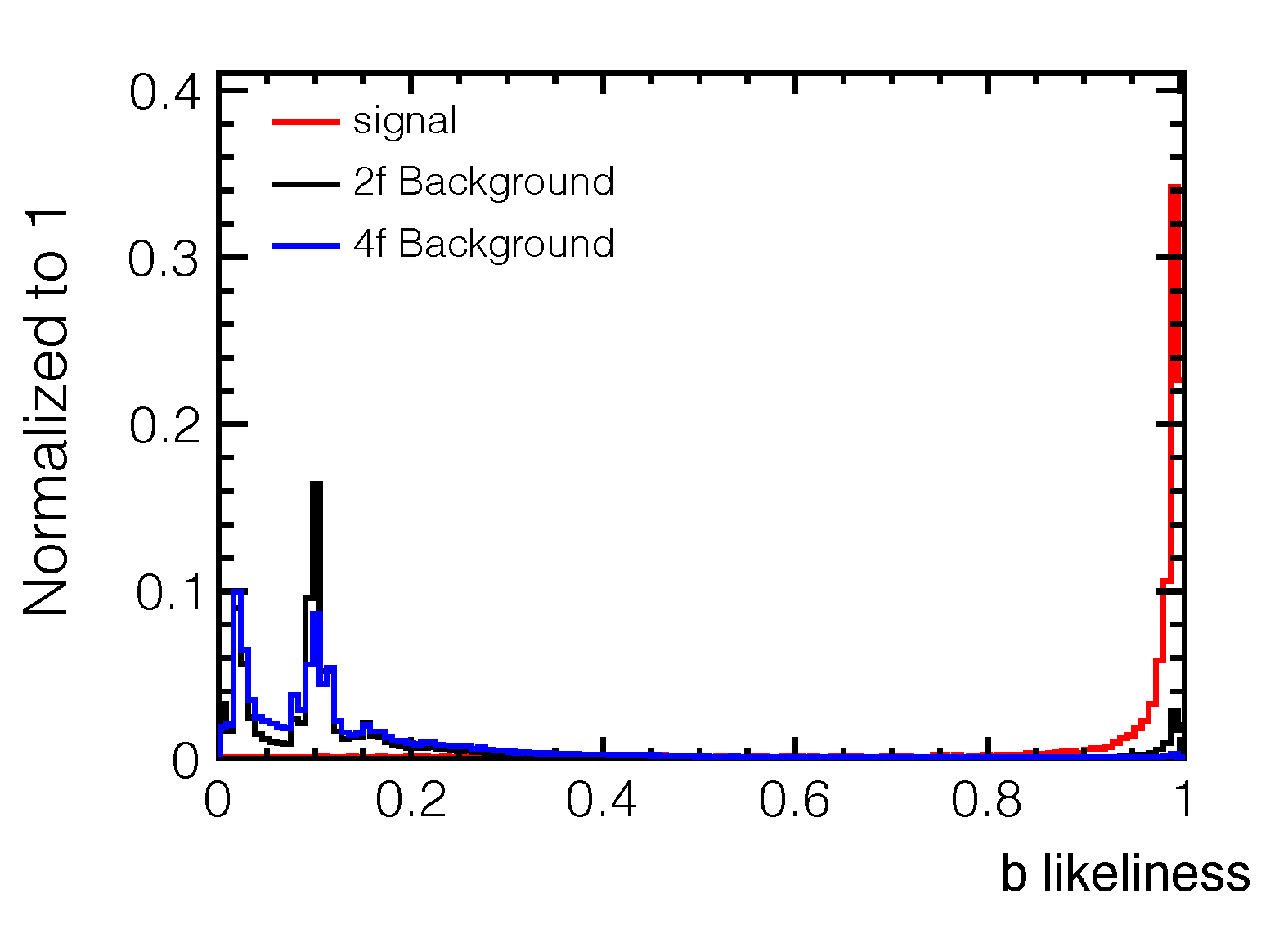}
	\end{flushleft}
	\end{minipage}
	\begin{minipage}{0.5\hsize}
	\begin{flushleft}
	\includegraphics[width=83mm,clip]{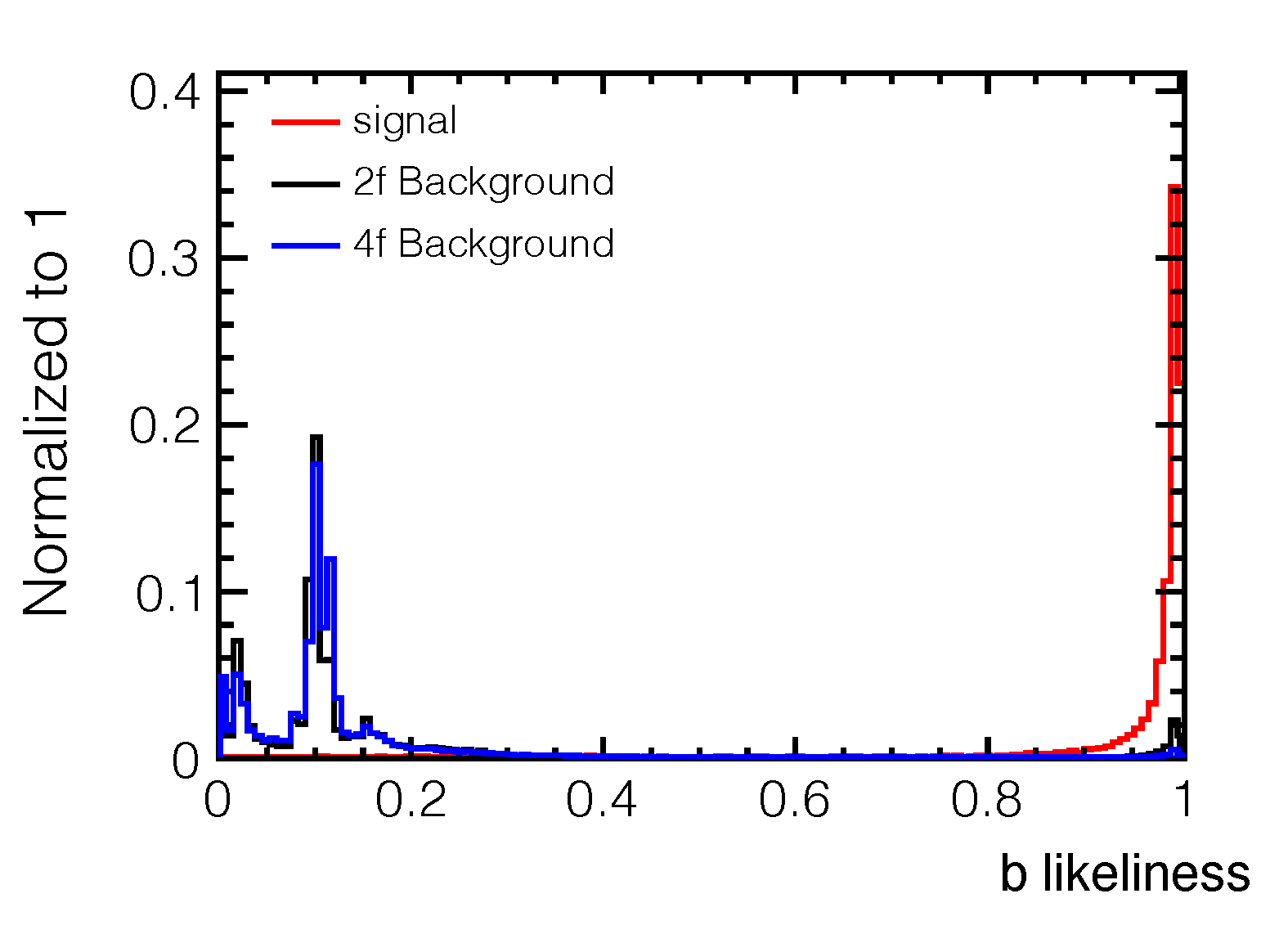}
	\end{flushleft}
	\end{minipage}
	\end{tabular}
\end{center}
\vspace{-6mm}
\caption{
\label{fig:bmax}
The distributions of the larger $b$-tag probability of a jet of signal, $2f$ and $4f$ events for the left-handed case (left plot) and the right-handed case (right plot).} 
\end{figure}

\begin{figure}[htbp]
\begin{center}
	\begin{tabular}{c}
	\begin{minipage}{0.5\hsize}
	\begin{flushleft}
	\includegraphics[width=83mm,clip]{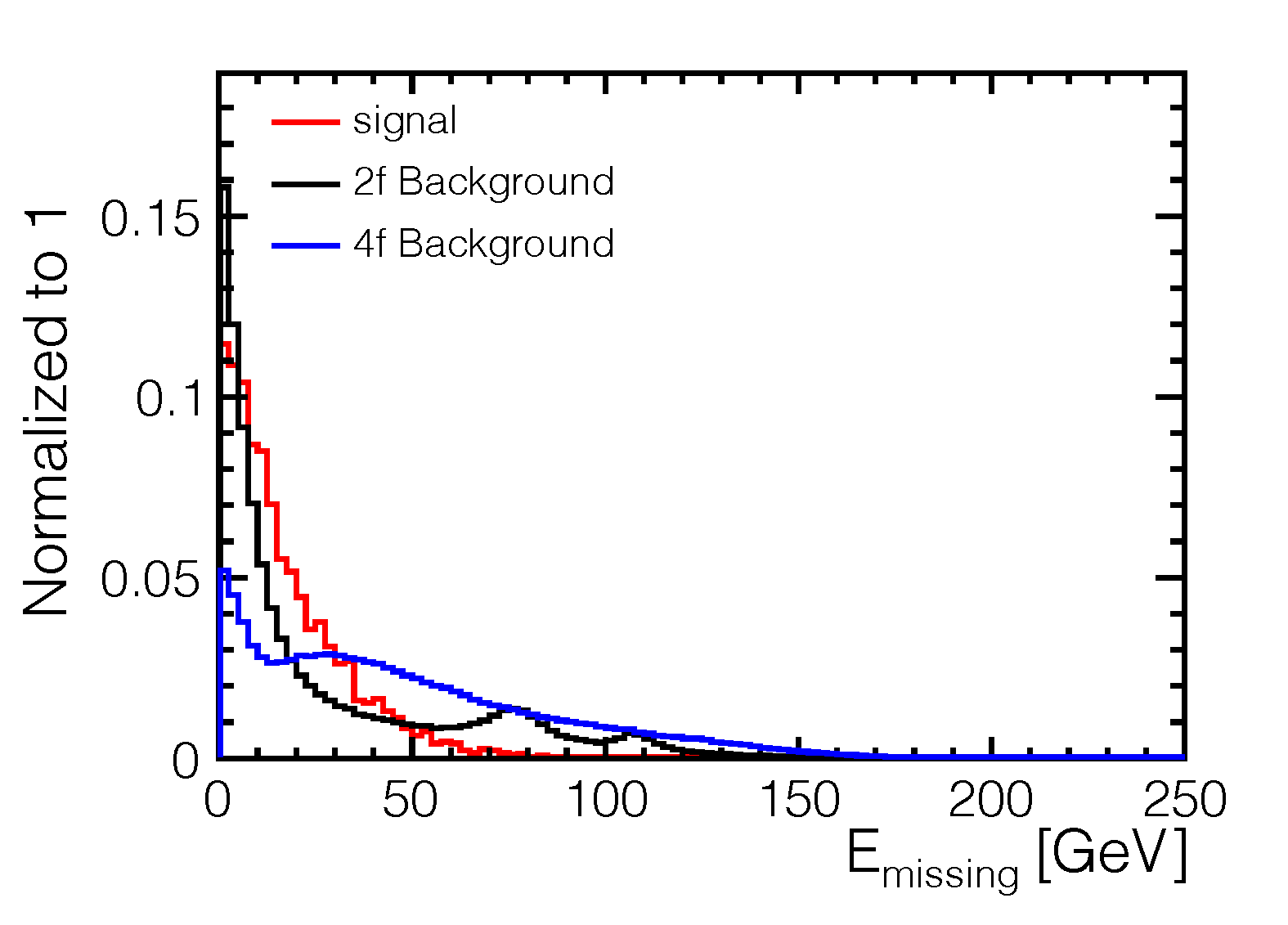}
	\end{flushleft}
	\end{minipage}
	\begin{minipage}{0.5\hsize}
	\begin{flushleft}
	\includegraphics[width=83mm,clip]{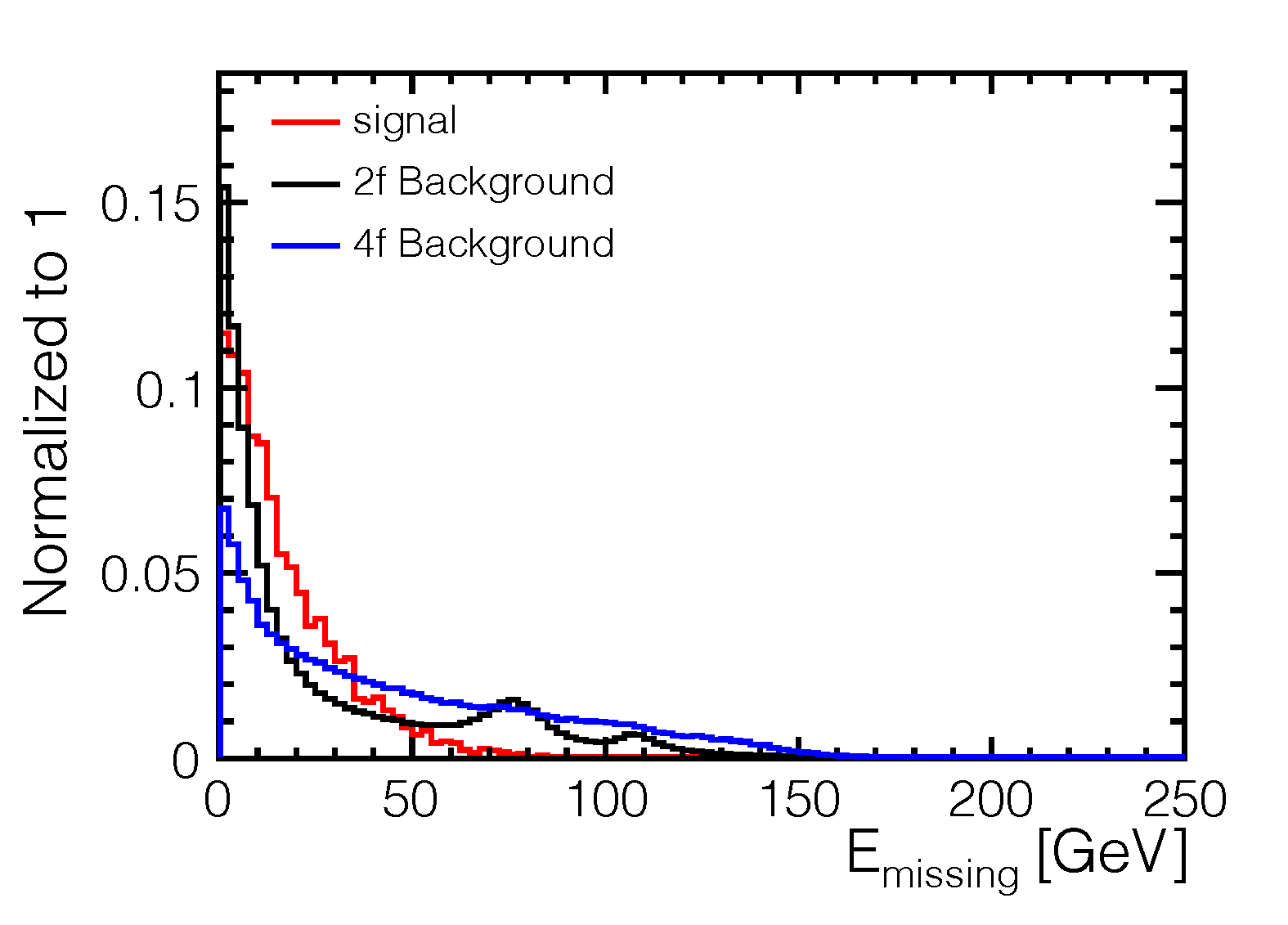}
	\end{flushleft}
	\end{minipage}
	\end{tabular}
\end{center}
\vspace{-6mm}
\caption{
\label{fig:emis}
The distributions of the total missing energy of signal, $2f$, and $4f$ events for the left-handed case (left plot) and the right-handed case (right plot).} 
\end{figure}

For each event, the highest $b$-tag probability of the two jets is required to greater than 0.77. The total missing energy is required to be less than 35 GeV. The remaining background events after these cuts are supplied to a BDT multivariate analysis (MVA), in which the input variables are the energy of the signal photon, the invariant mass of two b-jets, the angles between the signal photon and each of the two b-jets, the angle between the two b-jets. With the TMVA package~\cite{TMVA} of ROOT 5 MVA is first trained using half of the events and then is applied to the other half. The MVA output value for each event is required to be larger than 0.025. As a final event selection cut, the polar angle of the signal photon is required to be $|\cos\theta_{h \gamma} |<0.92$. The dominant background events after all selection cuts are from $e^+e^-\to\gamma Z^* \to \gamma b\bar{b}$ which are essential irreducible. \Figref{fig:bmax} and \Figref{fig:emis} show the distributions of the larger $b$-tag probability of a jet and total missing energy per event respectively for both left- and right-handed cases.

\subsection{$h\to WW^*$ semi-leptonic channel}
\begin{figure}[htbp]
\begin{center}
	\begin{tabular}{c}
	\begin{minipage}{0.5\hsize}
	\begin{flushleft}
	\includegraphics[width=83mm,clip]{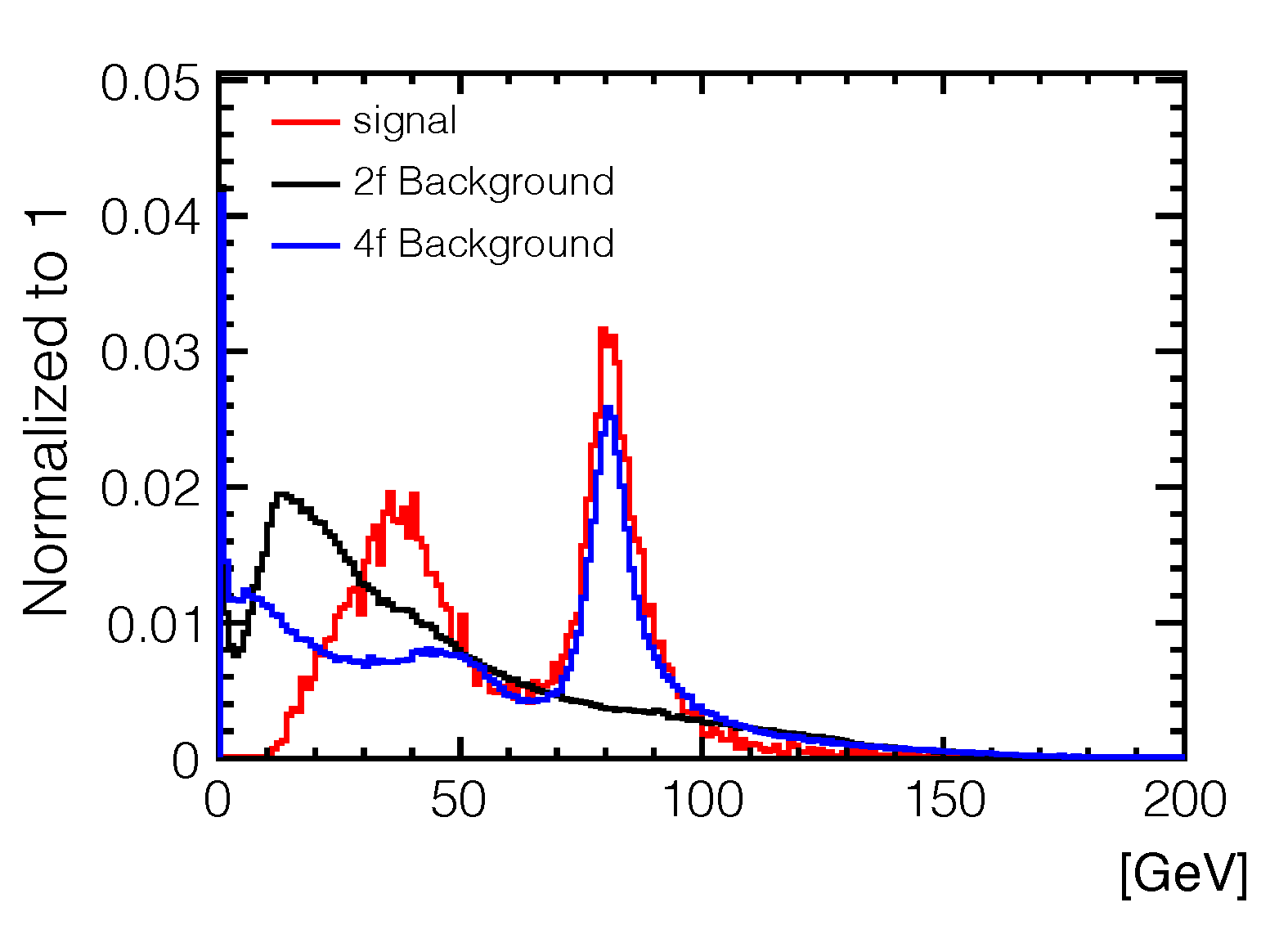}
	\end{flushleft}
	\end{minipage}
	\begin{minipage}{0.5\hsize}
	\begin{flushleft}
	\includegraphics[width=83mm,clip]{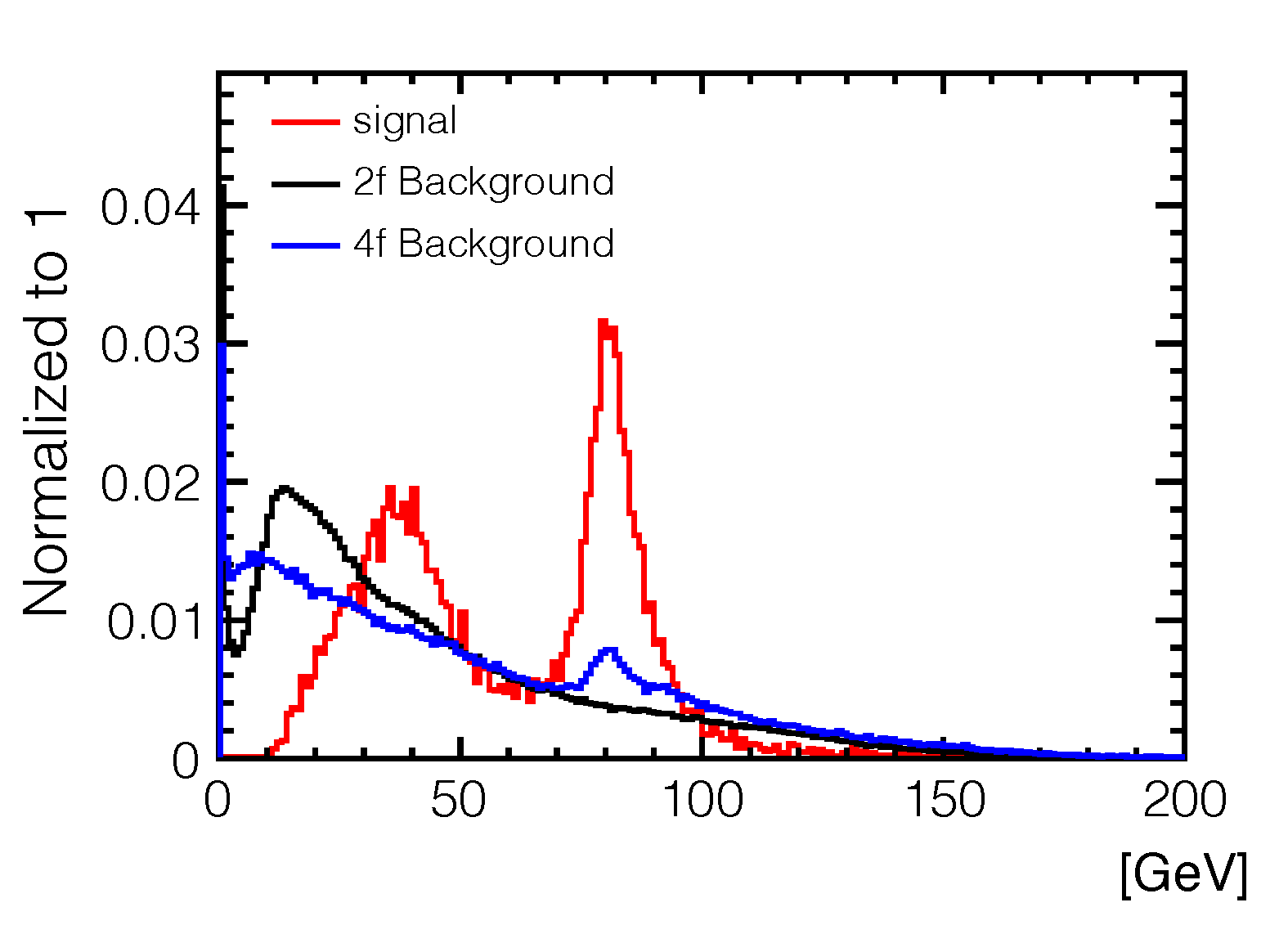}
	\end{flushleft}
	\end{minipage}
	\end{tabular}
\end{center}
\vspace{-6mm}
\caption{
\label{fig:mw1} 
Left (Right) plot shows the distributions of teh invariant mass of reconstructed $W_1$ of the signal and background events for the left- (right-) handed polarisation.} 
\end{figure}

\begin{figure}[htbp]
\begin{center}
	\begin{tabular}{c}
	\begin{minipage}{0.5\hsize}
	\begin{flushleft}
	\includegraphics[width=83mm,clip]{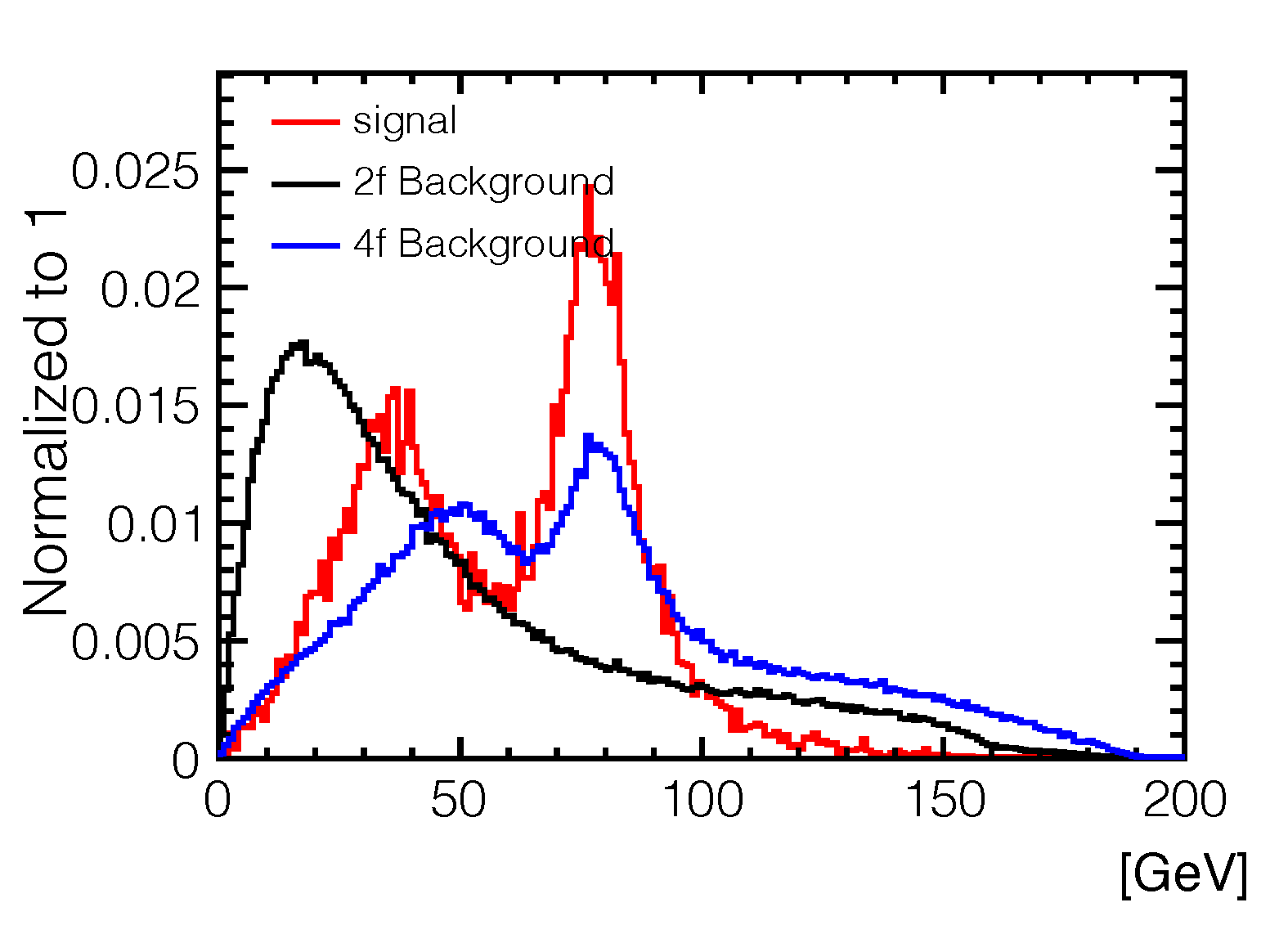}
	\end{flushleft}
	\end{minipage}
	\begin{minipage}{0.5\hsize}
	\begin{flushleft}
	\includegraphics[width=83mm,clip]{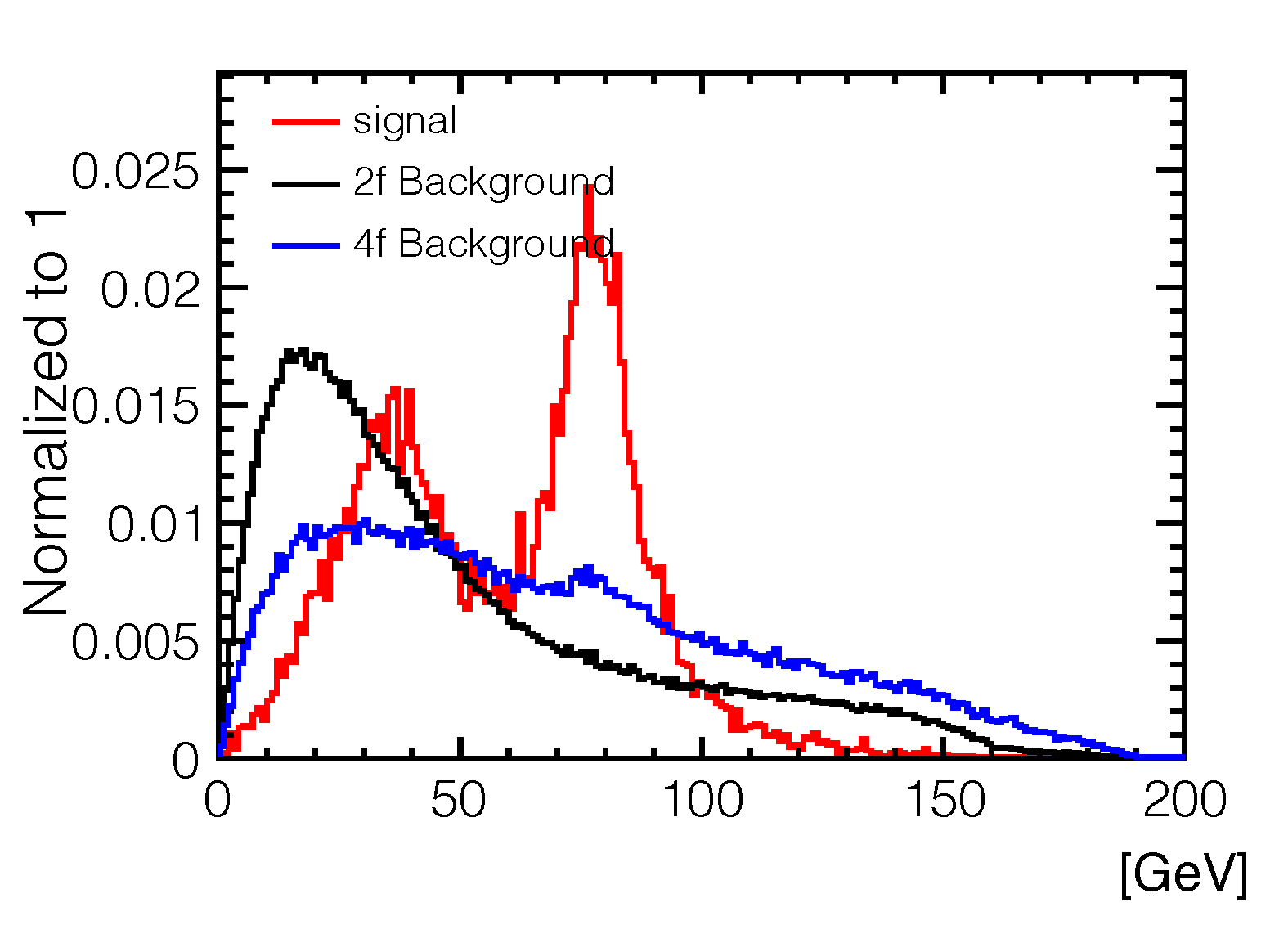}
	\end{flushleft}
	\end{minipage}
	\end{tabular}
\end{center}
\vspace{-6mm}
\caption{
\label{fig:mw2}
Left (right) plot shows the distributions of the invariant mass of reconstructed $W_2$ of the signal and background events for the left- (right-) handed polarisation.} 
\end{figure}

The signal events are characterized as having one monochromatic energetic photon with an energy of 93 GeV, one isolated electron or muon (we didn't study $\tau$ signal channel), two jets that are not b-tagged, and large missing energy from a neutrino accompanying lepton in leptonic W decay. After the signal photon is identified as above, an isolated lepton algorithm is applied to find the signal electron or muon using IsolatedLeptonTagging~\cite{ILT} implemented in iLCSoft. The remained particles are clustered into 2 jets using the Durham jet clustering algorithm, each of which is flavor tagged using LCFI+. The number of charged particles in each event is required to be at least three. The larger b-likeness of a jet is required to be less than 0.77 to extract $h \to b \bar{b}$ events. The two jets are paired as one signal $W$ boson (namely $W_1$), the isolated lepton and missing four-momentum are paired as the other signal $W$ boson (namely $W_2$). The two $W$ bosons should be one on-shell, invariant mass of which is required to be greater than 70.4 (71) GeV and less than 90.4 (89.8) GeV if the on-shell $W$ is $W_1$ ($W_2$), and one off-shell. The invariant mass of the two $W$ should be consistent with the Higgs mass. \Figref{fig:mw1} and \Figref{fig:mw2} show the distributions of $m_{W_1}$ and $m_{W_2}$ for the signal and background events for both polarization. The domimant background events are from $e^+e^-\to W^+W^-$ where one of the $W$ bosons is on-shell and one energetic photon is radiated from either the initial states or the final states.

\clearpage
\section{Result}
\subsection{Result for each channel and polarisation}
\Tabref{tbl:2} shows the number of signal and background events as well as signal significance in the $h \to b \bar{b}$ channel (left-handed case) after every event selection cut. The signal significance is defined as follows, 
\begin{eqnarray}
\text{Significance} = \frac{N_{s}}{\sqrt{N_{s}+N_{B}}},
\end{eqnarray}
where $N_{s}$ is the number of selected signal events and $N_{B}$ is the number of selected background events.

After all cuts, we expect 29 signal events and 12 thousand background events for the SM signal process, with a signal significance of 0.26$\sigma$. 
The 95$\%$ confidence level upper limit for the cross-section of $e^+e^- \to h \gamma $ with left-handed beam polarisation is obtained using \Equref{equ:sig}: $\sigma_{h\gamma}^L<2.6$ fb.

\begin{table}[htbp]
\begin{center}
\caption{The summary table of the number of events of $h \to b\bar{b}$ channel (left-handed case).}
\label{tbl:2} 
\begin{tabular}
{lccc}\hline 
  & \text { Total bg } & \text { Signal } & \text { Significance } \\ \hline 
  \text { Expected } & 1.4 $\times 10^{8}$ & 107 & 0.01 \\ 
  \text { Preselection } & 2.9 $\times 10^{7}$ & 100 & 0.02 \\ 
  \text { $b$-likeliness} $>$ 0.77 & 2.2 $\times 10^{6}$ & 90 & 0.06 \\
  $E_{mis} <$ 35 GeV & 1.9 $\times 10^{6}$ & 82 & 0.06 \\ 
  \text {  MVA cut} $>$ 0.025 & 19583 & 34 & 0.24 \\ -0.92 $< \cos \theta_{h \gamma} < $0.92 & 12422 & 29 & 0.26 \\ \hline
\end{tabular}
\end{center}
\end{table}

\begin{eqnarray}
\label{equ:sig}
\sigma_{h \gamma }^L&=&\sigma_{S M}+\frac{1.64}{\text { significance }} \sigma_{S M}\\
&=&2.6~\text {fb}
\end{eqnarray}

\Tabref{tbl:sig2}, \Tabref{tbl:4}, and \Tabref{tbl:5} is similar table for $h \to b\bar{b}$ channel (Right-Handed Case), $h \to WW^*$ semi-leptonic channel (Left-Handed Case), and $h \to WW^*$ semi-leptonic channel (Right-Handed Case). We summarize the 95$\%$ confidence level upper limit for the cross-section of $e^+e^- \to h \gamma $ in \Tabref{upperlimit}.

\begin{table}[htbp]
\begin{center}
\caption{The reduction table of the $h \to b\bar{b}$ channel (right-handed case).}
\label{tbl:sig2} 
\begin{tabular}
{lccc}\hline & 
\text { total bg } & \text { Signal } & \text { Significance } \\ \hline 
\text { Expected } & 7.8 $\times 10^{7}$ & 11.2 & 0.001 \\ 
\text { Preselection } & 2.3 $\times 10^{7}$ & 10.3 & 0.002 \\ 
$b$-likeliness $>$ 0.77 & 1.5 $\times 10^{6}$ & 9.4 & 0.008 \\ 
$E_{mis }<$35 GeV & 1.3 $\times 10^{6}$ & 8.4 & 0.007 \\ 
\text {  MVA cut} $>$ 0.025 & 1.0 $\times 10^{4}$ & 3.4 & 0.034 \\ 
$-0.92 < \cos \theta_{h \gamma } < 0.92$ & 5.9 $\times 10^{3}$ & 3.0 & 0.039 \\ \hline
\end{tabular}
\end{center}
\end{table}

\begin{table}[htbp]
\begin{center}
\caption{The summary table of the $h \to WW^*$ semi-leptonic channel (left-handed case).}
\label{tbl:4} 
\begin{tabular}{lccc}\hline & \text { total bg } & \text { Signal } & \text { Significance } \\ \hline \text { Expected } & 1.4 $\times 10^{8}$ & 18.0 & 0.003 \\
\text { Preselection } & 1.3 $\times 10^{7}$ & 10.5 & 0.004 \\ 
 \text {No. of charged particle} $>$ 3 & 3.1 $\times 10^{5}$ & 5.4 & 0.010 \\ $|m_{W_1}-80.4$ GeV $| < 10$ GeV \text { or } & & & \\ 
$ | \mathrm{m_{W_2}} -80.4$ GV $| < 9.4$ GeV& 1.9 $\times 10^{5}$ & 3.7 & 0.009 \\
$b$-likeliness $<$ 0.77 & 1.8 $\times 10^{5}$ & 3.7 & 0.009 \\ 
 MVA cut $>$ 0.1 & 41 & 1.0 & 0.16 \\ 
$-0.93 < \cos \theta_{h \gamma } < 0.93$ & 8 & 0.9 & 0.31 \\ \hline
\end{tabular}
\end{center}
\end{table}

\begin{table}[htbp]
\begin{center}
\caption{The reduction table of the $h \to WW^*$ semi-leptonic channel (right-handed case).}
\label{tbl:5} 
\begin{tabular}{lccc}\hline & \text { total bg } & \text { Signal } & \text { Significance } \\\hline 
\text { Expected } & 7.8 $\times 10^{7} $& 1.9 & 0.000 \\ 
\text { Preselection } & 1.2 $\times 10^{7}$ & 2.0 & 0.000 \\ 
\text { \# of charged particle} $>$ 3 & 8.6 $\times 10^{4}$ & 1.5 & 0.002 \\ 
$|m_{W_1}-80.4$ GeV $| < 10$ GeV \text { or } & & & \\ $|m_{W2} -80.4$ GeV $| < 9.4$ GeV & 3.2 $\times 10^{4}$ & 0.4 & 0.002 \\ 
$b$-likeliness $<$ 0.77 & 2.6 $\times 10^{5}$ & 0.4 & 0.002 \\ \text {  MVA cut }$>$ 0.1 & 74 & 0.1 & 0.01 \\ 
-0.93 $< \cos{\theta_{h \gamma }} <$ 0.93 & 5 & 0.1 & 0.04 \\ \hline\end{tabular}
\end{center}
\end{table}

\begin{table}[htbp]
\begin{center}
\caption{The summary table of the 95 $\%$ confidence level upper limit for the cross-section of $e^+e^- \to h \gamma $.}
\label{upperlimit} 
\begin{tabular}
{lccc}\hline 
\text { Channel } & \text { Polarisation } & \text { Upper limit [fb] } \\ 
\hline 
 \text { $h \to b\bar{b}$ } & Left-handed  & $\sigma_{h\gamma}^L<2.6$  \rule[0mm]{0mm}{5mm} \\ 
\text { $h \to b\bar{b}$ } & Right-handed &$\sigma_{h\gamma}^R<0.7$\\ 
\text { $h \to WW^*$ semi-leptonic channel}  & Left-handed  & $\sigma_{h\gamma}^L<2.2$ \\
\text { $h \to WW^*$ semi-leptonic channel} & Right-handed & $\sigma_{h\gamma}^R<0.7$ \\ \hline
\end{tabular}
\end{center}
\end{table}

\subsection{Combined result}
We combined results of  the two signal channels $h\to b \bar{b}$ and $h\to WW^*\to q\bar{q} l\nu (l=e,\mu)$. The results of 95$\%$ C.L. upper limit on cross section of $e^+e^- \to h \gamma$ are:
\begin{eqnarray}
\sigma_{h \gamma }^L &<& 1.8 ~\text{fb} \\
\sigma_{h \gamma }^R &<& 0.5 ~\text{fb}. 
\label{Equ:10}
\end{eqnarray}

\clearpage
\section{Uncertainty due to finite MC statistics (Left-handed)}

Due to the limited number of MC events available, we tried to estimate the uncertainty of the count of remained signal and background events after all cuts. The uncertainty due to the signal MC statistics is negligible. The uncertainty due to the MC statistics of dominant background in each channel can be rather large, simply because the weights of those events are as high as between 20 to 40. The large uncertainty mainly appears after the tight MVA cut. Thus we tried to re-estimate the efficiency of MVA cut by loosening the cuts applied before MVA cut to gain statistics, which however is in principle only accurate when the MVA cut is uncorrelated to the other cuts. We then take the difference between this re-estimated efficiency and the nominal efficiency (aka those in the result sections) as a conservative estimation of uncertainty due to MC statistics, which are reported in \Tabref{tb:re-estimated} and \Tabref{tbl:7}, available only for left-handed beam polarisation for two signal channels.

\begin{table}[htbp]
\begin{center}
\caption{Estimation of uncertainty of the number of signal and background events due to limited MC statistics for $h\to b\bar{b}$ channel with left-handed beam polarisation.}
\label{tb:re-estimated} 
\begin{tabular}{ccccc}\hline 
& \text { total bg } & \text { Signal } & \text { Significance } & 95 \% \text { C.L. upper limit } \\ 
\text { on $\sigma_{h \gamma }$ (fb) } \\ \hline 
\text { Nominal } & 12422 & 29 & 0.29 & 2.6 \\ \text { Conservative } & 13488 & 29 & 0.25 & 2.7 \\ \hline
\end{tabular}
\end{center}
\end{table}

\begin{table}[htbp]
\begin{center}
\caption{Estimation of uncertainty of the number of signal and background events due to limited MC statistics for $h\to WW^*$ semi-leptonic channel with left-handed beam polarisation.}
\label{tbl:7} 
\begin{tabular}{ccccc}\hline & \text { total bg } & \text { Signal } & \text { Significance } & 95 \% \text { C.L. upper limit } \\ \text { on $\sigma_{h \gamma }$ (fb) }\\ \hline \text { Nominal } & 8 & 0.9 & 0.31 & 2.2 \\ \text { Conservative } & 92 & 0.9 & 0.09 & 6.5 \\ \hline\end{tabular}
\end{center}
\end{table}

\section{Summary and Future Study}
We have performed a full simulation study of $e^+e^- \to h \gamma $ at 250 GeV ILC using the ILD detector. The significance of the SM signal event with 900 fb$^{-1}$ is found to be 0.40$\sigma$ for the left-handed case and of 0.06$\sigma$ for the right-handed case. The constraints on new physics effects in this process are reported as the 95\% C.L. upper limit of the $h \gamma$ production cross-section which found to be $\sigma_{h\gamma}^L (\sigma_{h\gamma}^R) <$ 1.8 (0.5) fb. We also estimated uncertainty due to finite MC statistics for left-handed case.
As a next step, we plan to investigate constraints that can be set on the couplings $\zeta_{AZ}$ and $\zeta_A$, as well as on model parameters in concrete BSM models. We will put an effort in the interpretation of the obtained results in the context of concrete BSM models. In addition, we'll try to understand the implications of the obtained results on a global EFT fit of the Higgs couplings.

\section*{Acknowledgements}
We would like to thank the LCC generator working group and the ILD software working group for providing the simulation and reconstruction tools and producing the Monte Carlo samples used in this study.
This work has benefited from computing services provided by the ILC Virtual Organization, supported by the national resource providers of the EGI Federation and the Open Science GRID.

\end{document}